\documentclass[11pt]{article}

\usepackage{amsmath,amssymb}

\usepackage{epsfig}
\usepackage{graphicx}               
\usepackage{url}

\numberwithin{equation}{section}

\newcommand{\nc}{\newcommand}

\nc{\beq}{\begin{equation}}
\nc{\eeq}{\end{equation}}
\nc{\beqa}{\begin{eqnarray}}
\nc{\eeqa}{\end{eqnarray}}
\nc{\bea}{\begin{eqnarray}}
\nc{\eea}{\end{eqnarray}}
\nc{\ra}{\rightarrow}
\nc{\lsim}{\begin{array}{c}\,\sim\vspace{-21pt}\\< \end{array}}
\nc{\gsim}{\begin{array}{c}\sim\vspace{-21pt}\\> \end{array}}
\nc{\Tr}{{\rm Tr}}
\nc{\tr}{{\rm tr}}
\nc{\slsh}{\slash\hspace*{-0.22cm}}

\def\be{\begin{equation}}
\def\ee{\end{equation}}
\def\bea{\begin{eqnarray}}
\def\eea{\end{eqnarray}}
\def\bit{\begin{itemize}}
\def\eit{\end{itemize}}

\def\to{\rightarrow}

\title{
\vspace*{-2.3cm}
\begin{flushright}
\normalsize{
SLAC-PUB-15255, SU-ITP-12/28
  }
\end{flushright}
\vspace{1.5cm}
\Large
\textbf{
Large N (=3) Neutrinos and Random Matrix Theory
}\vspace*{1.0cm}
}

\author{Yang Bai$^{a,b}$ and Gonzalo Torroba$^{b,c}$
\vspace{5mm}
\\
$^{a}$  \normalsize\emph{Department of Physics, University of Wisconsin, Madison, WI 53706, USA}  
 \vspace{1mm} \\ 
$^{b}$ \normalsize\emph{SLAC National Accelerator Laboratory, 2575 Sand Hill Road, Menlo Park, CA 94025, USA} 
 \vspace{1mm} \\ 
 $^{c}$ \normalsize\emph{
Stanford Institute for Theoretical Physics, Stanford University, Stanford, CA 94305, USA  } 
}

\date{}

\begin{document}
\setcounter{page}{0}
\maketitle

\vspace*{1cm}
\begin{abstract}
The large N limit has been successfully applied to QCD, leading to qualitatively correct results even for $N=3$. In this work, we propose to treat the number $N=3$ of Standard Model generations as a large number. Specifically, we apply this idea to the neutrino anarchy scenario and study neutrino physics using Random Matrix Theory, finding new results in both areas. For neutrino physics, we obtain predictions for the masses and mixing angles as a function of the generation number $N$. The Seesaw mechanism produces a hierarchy of order $1/N^3$ between the lightest and heaviest neutrino, and a $\theta_{13}$ mixing angle of order $1/N$, in parametric agreement with experimental data when $N$ goes to $3$. For Random Matrix Theory, this motivates the introduction of a new type of ensemble of random matrices, the ``Seesaw ensemble." Basic properties of such matrices are studied, including the eigenvalue density and the interpretation as a Coulomb gas system. Besides its mathematical interest, the Seesaw ensemble may be useful in random systems where two hierarchical scales exist. 
\end{abstract}

\thispagestyle{empty}

\bigskip
\newpage

\baselineskip18pt

\tableofcontents

\vskip 1cm

\section{Introduction}
\label{sec:introduction}

The large $N$ limit is a very useful tool in various theoretical models, both at a qualitative and a quantitative level~\cite{Coleman}. Historically, one of the most important examples came from 't Hooft's large $N$ analysis of QCD~\cite{'tHooft:1973jz}. Although the number of colors $N_c=3$ is not very large, many features of mesons and baryons calculated in a $1/N_c=1/3$ expansion are in good agreement with experiment~\cite{Manohar:1998xv}. One cannot help but notice that in the Standard Model (SM), the number ``3'' also appears as the number of fermion generations. Here, we propose to treat the number of families $N_f=3$ in a large $N$ approximation, and study the properties of fermion masses and mixings as a function of the expansion parameter $1/N_f = 1/3$. 

In this paper, we make the first attempt to use the large $N$ limit  to understand some part of the flavor problem (i.e. the origin of masses and mixings of the SM fermions). More specifically, we will apply this idea to the SM neutrinos assuming the mass anarchy scenario of~\cite{Hall:1999sn,Haba:2000be}, where the neutrino couplings are treated as random variables. In light of the recent measurement of $\theta_{13}$ from Daya Bay~\cite{An:2012eh} and later  confirmed by Reno~\cite{Ahn:2012nd}, the anarchy hypothesis is consistent with the measured value~\cite{deGouvea:2003xe, deGouvea:2012ac}. We will not consider the quark sector here (which requires additional structure), although we will comment briefly on possible large $N$ applications in \S \ref{sec:conclusion}.

Extending the $3\times 3$ random matrix for the three active neutrinos to an $N\times N$ random matrix (hereafter $N=N_f$ denotes the number of generations), we have to analyze the statistical properties of the neutrino spectrum and mixing matrix. This is a classic problem in Random Matrix Theory (RMT); see e.g.~\cite{mehta, Stephanov:2005ff} for reviews. RMT has an incredible range of applications, from medical HIV studies~\cite{HIV} to the string landscape~\cite{Denef:2004cf, Marsh:2011aa, Chen:2011ac}, and now we are adding neutrino physics to this list. 

RMT provides analytic results for the correlation functions of eigenvalues of large rank matrices. For example, the eigenvalues of Hermitian complex matrices obey the well-known Wigner semicircle distribution. Knowing the spectral density for the neutrino masses would already provide valuable information. It will allow us to calculate the expectation value of the ratio of the lightest and heaviest masses, which will in turn determine how hierarchical the mass spectrum is. Although experimentalists have measured all three mixing angles of the Pontecorvo-Maki-Nakagawa-Sakata (PMNS) matrix, it is still not clear whether the neutrino mass spectrum is normal or inverted, and hierarchical or quasi-degenerate. Our analysis for general $N$ will reveal that the statistical properties of the spectrum depend sensitively on the underlying mass mechanism.

There exist different possibilities for the neutrino mass matrices. Our main focus will be on the Seesaw mechanism, which provides a dynamical explanation for the smallness of neutrino masses. We will demonstrate that the Seesaw mechanism is not only required to explain the hierarchy between the neutrino mass scale and the weak scale, but also preferred by the large $N$ behavior of our RMT analysis and the current experimental results on the neutrino mass-squared differences. Besides providing a useful tool to analyze the spectrum, the connection with RMT will also lead us to define a new type of random matrix ensemble, as we describe in more detail below.

We have attempted to make our paper accessible to both the neutrino physics and RMT communities. We begin in \S\ref{sec:RMT} with a short review of the results in RMT that we will need, including the Gaussian unitary ensemble in \S\ref{subsec:gaussian} and the Coulomb gas picture in \S\ref{subsec:coulomb}. Next, we apply RMT techniques to the simplest case of complex Majorana neutrinos in \S\ref{sec:Majorana}. Our main results are presented in \S\ref{sec:see-saw}, where we analyze the large $N$ spectrum of random Seesaw neutrinos, and also define and study the new ``Seesaw ensemble.'' In \S\ref{sec:application} we apply these results to the phenomenologically relevant $N=3$ case and compare with experimental data. Our conclusions and future directions are summarized in \S\ref{sec:conclusion}.

\section{Review of random matrix theory}
\label{sec:RMT}

Let us begin by reviewing some basic results from random matrix theory that will be needed in the following sections. For a pedagogical exposition, we refer the reader to~\cite{mehta, Stephanov:2005ff}.

\subsection{Gaussian unitary ensemble}
\label{subsec:gaussian}

Random matrix theory is the study of the statistical properties of eigenvalues of very large matrices. We will first consider $N \times N$ Hermitian complex matrices $M= A + A^\dag$, where each element of $A$ is an independently distributed random variable. The space of such matrices is known as the Gaussian unitary ensemble (GUE). 

Writing the probability distribution as $P(M) dM$, the defining properties of the GUE are that: a) it is invariant under unitary transformations, $P(M) dM = P(M') dM'$, with $M^\prime= U^\dag M U$; and b) the matrix elements are statistically independent, namely $P(M) = \prod_{i \le j} P_{ij}(M_{ij})$. From this, it follows that~\cite{mehta}
\beq
P(M) dM \propto \,dM\, e^{- \frac{N}{2} \,\tr[M^\dag M]}\,,
\eeq
where $M$ has been linearly redefined to set the center at zero and to fix the coefficient in the exponential. The matrix $M$ is diagonalized by
\beq
M = U \Lambda U^\dag \,.
\eeq
Here, $\Lambda=\text{diag}(m_1,\ldots,m_N)$ with $m_i$ as the eigenvalues of the matrix with the convention $m_i \leq m_j$ for $i<j$ and $U$ a special unitary matrix with $N(N-1)/2$ independent parameters. Changing variables to $\Lambda$ and $U$ then gives
\beq\label{eq:distrib1}
P(M) dM = C_N\,dU\, \prod_i dm_i\, \prod_{i<j} (m_i-m_j)^2\,e^{-\frac{N}{2} \sum_{i}m_i^2}\,.
\eeq
Here, $dU$ is the invariant Haar measure for $U(N)$ and $C_N$ is a normalization constant.

We see that at large $N$ the distribution of angles and phases contained in $U$ becomes flat in the natural group theory coordinates. This turns out to have important consequences for the neutrino mixing angles, analyzed by~\cite{Hall:1999sn, Haba:2000be}. On the other hand, the eigenvalues have nontrivial statistical properties. These are characterized by the k-point correlation functions
\beq
R_k(m_1,\ldots,m_k)= \frac{N!}{(N-k)!} \int dm_{k+1} \ldots dm_N\,P(m_1,\ldots,m_N)\,.
\eeq
Of particular interest for us will be the one-point function
\be\label{eq:rhodef}
\left\langle \rho(m) \right\rangle \equiv \left\langle \sum_i \delta(m-m_i) \right\rangle=N \int dm_2 \ldots dm_N \,P(m,m_2,\ldots,m_N)\,.   
\ee
This gives the eigenvalue density, and its inverse $1/\rho(m)$ determines the mean level spacing around $m$.

The GUE can be solved exactly in the limit $N \to \infty$. This is done by recognizing that the distribution in Eq.~(\ref{eq:distrib1}) can be written as a product of orthogonal Hermite polynomials. The correlators are then of the form
\beq\label{eq:Rk}
R_k(x_1,\ldots,x_k) = \det [ K_N(x_i,x_j)]_{i,j=1,\ldots,k}\,,
\eeq
where the kernel $K_N$ is defined as
\beq
K_N(x,y) \equiv e^{-(x^2+y^2)/2}\sum_{n=0}^{N-1} H_n(x) H_n(y) \,.
\eeq
Here we have introduced the rescaled variables
\beq\label{eq:x-rescaled}
x_k \equiv \sqrt{\frac{N}{2}}m_k\,.
\eeq

This result implies that at large $N$ the eigenvalue density is given by the Wigner semicircle distribution,
\beq\label{eq:semicircle}
\renewcommand{\arraystretch}{1.5}
\left\langle \rho(x) \right\rangle \approx 
\left\lbrace
\begin{matrix}
\frac{1}{\pi} \sqrt{2N-x^2} & \text{for}\;x^2 \leq 2N \,, \\
0 & \text{for}\;x^2 > 2N \,.
\end{matrix} \right.
\eeq
Changing back to the original eigenvalues $m$ yields $\rho(m)= \frac{N}{2\pi} \sqrt{4-m^2}$ for $m^2<4$ and zero otherwise. Thus, the spacing of eigenvalues near the origin is of order $1/N$. Of course, the range of the semicircle ($4$ in this case) is conventional, being related to the normalization of the exponential in Eq.~(\ref{eq:distrib1}).

\subsection{The Coulomb gas picture and generalizations}
\label{subsec:coulomb}

We will now review the physical interpretation of the GUE as a Coulomb gas, introduced by Dyson~\cite{Dyson:1962es,Dyson2}. This picture will be useful when we study the ``Seesaw ensemble'' in \S\ref{sec:see-saw}.

The basic idea is that Eq.~(\ref{eq:distrib1}) can be interpreted as the thermal partition function for $N$ charged particles with positions $(x_1, \ldots , x_N)$ moving on a fixed line in two dimensions, under the influence of a harmonic potential and electrostatic repulsion. In more detail, the Hamiltonian for this system is
\beq\label{eq:H1}
H =  \sum_i x_i^2 - \sum_{i<j} \log (x_i-x_j)^2\,.
\eeq
Then the partition function
\beq
Z_N (\beta) = Z_0 \int\,\prod_i dx_i\,e^{-\beta H}
\eeq
reproduces the GUE distribution if the inverse temperature is fixed at $\beta=1$. [These expressions are given in terms of the rescaled variables in Eq.~(\ref{eq:x-rescaled}).]

Physically, the properties of the eigenvalue distribution appear from a competition between the harmonic potential and the electrostatic repulsion. If the spacing between charged particles decreases with $N$, at large $N$ the system admits a continuum approximation 
\beq\label{eq:Hcont}
H=  \int dx\,\rho(x) x^2 - \int dx dy\,\rho(x) \rho(y) \log |x-y| \,,
\eeq
where $\rho(x)$ is the level density introduced in Eq.~(\ref{eq:rhodef}).\footnote{The factor of $2$ from $\log(x-y)^2$ was canceled against a factor of $1/2$ from the requirement $i<j$ in Eq.~(\ref{eq:H1}).} The density is determined by extremizing $H$, subject to the constraints $\rho(x) \ge 0$ and $\int dx \rho(x)= N$. The stationary condition leads to the integral equation
\beq
- x^2 + \int dy\,\rho(y) \log|x-y|= \text{const}\,.
\eeq
The solution to this equation is given by the semicircle distribution Eq.~(\ref{eq:semicircle}).

This picture suggests a way of understanding non-Gaussian ensembles, with probability distribution
\beq
P(M) \propto e^{-\tr\left[V(M,\,M^\dag)\right]}\,.
\eeq
Such an ensemble is equivalent to a system of charged particles with a Hamiltonian
\beq\label{eq:Hgeneral}
H= \sum_i V(x_i) - \sum_{i < j}\,\log(x_i-x_j)^2\,.
\eeq
The particles interact with each other via the logarithmic Coulomb repulsion, and moreover each of them is subjected to a nontrivial confining potential $V(x)$. The level density $\rho(x)$ of the ensemble can be obtained, as before, by going to a continuum limit at large $N$ and extremizing the analog of Eq.~(\ref{eq:Hcont}) with $x^2$ replaced by $V(x)$. The result is an integral equation for $\rho(x)$,
\beq\label{eq:Vrho}
- V(x) + \int dy\,\rho(y) \log|x-y|= \text{const}\,.
\eeq
This relation will be applied in \S\ref{sec:see-saw}, where knowledge of the mass distribution generated by the Seesaw mechanism will be used to construct a suitable confining potential $V(x)$.

\section{Majorana neutrinos and $1/N$ hierarchies}
\label{sec:Majorana}

It is helpful to begin our analysis with the simplest case of neutrinos with random Majorana masses. Results for Dirac neutrinos are qualitatively similar. We consider $N$ neutrinos $\nu_i$ with Majorana masses
\beq
{\cal L}_m = - \frac{1}{2}M_\nu^{ij} \nu^T_i\,{\cal C}\, \nu_j + h.c.\,,
\eeq
where ${\cal C}$ is the charge conjugation operator. $M_\nu$ is an $N \times N$ complex symmetric matrix. For the SM, $N=3$ and ${\cal L}_m$ comes from a dimension-five operator $(\tilde{H} L)^2$. As in~\cite{Hall:1999sn, Haba:2000be}, $M_\nu$ is taken to be random. Instead of specializing from the start to $N=3$, we will determine the properties of neutrino masses at large $N$ and then take $N \to 3$. At large $N$ the lightest neutrino will be found to have a mass suppressed by $1/N$. The results from this section will also help to develop intuition and tools to study in \S\ref{sec:see-saw} the more interesting and nontrivial random Seesaw mechanism case.

We first analyze the statistical properties of random  complex symmetric Majorana mass matrices. In order to find the probability distribution of eigenvalues, we perform a singular value decomposition
\beq\label{eq:Mnu-SVD}
M_\nu = U \Lambda U^T\;,\quad \;\Lambda=\text{diag}(m_1,\ldots,m_N)\;, \quad \;U \in U(N)\,.
\eeq
The measure becomes~\cite{Haba:2000be}
\beq
dM_\nu = dU \,\prod_k m_k dm_k\, \prod_{i<j} \left|m_i^2 -m_j^2\right|\,.
\eeq
As reviewed in \S\ref{sec:RMT}, statistical independence of the elements $(M_\nu)_{ij}$ and invariance under unitary transformations implies that the probability distribution is Gaussian. Therefore, the complex Majorana neutrino masses are distributed according to
\beq
P(M_\nu) dM_\nu = C_N\, dU \,\prod_k m_k dm_k\,\prod_{i<j} \left|m_i^2 -m_j^2\right|\, e^{-\frac{N}{4} \sum_i m_i^2}\,,
\eeq
which is known as the {\it Altland-Zirnbauer} CI ensemble~\cite{Altland:1997zz}. The eigenvalues occur in pairs $\pm m_i$, which will be important shortly. (This is also why the exponential contains an extra factor of $1/2$ as compared to the GUE.)

The statistical distribution of neutrino masses is equivalent to a Coulomb gas system with Hamiltonian
\beq
H =  \frac{1}{2} \sum_i x_i^2 - \sum_{i<j} \log \left| x_i^2-x_j^2\right| - \sum_i \log |x_i|\,,
\eeq
where $x_i=\sqrt{\frac{N}{2}} m_i$. The last term encodes an electrostatic repulsion between the images $\pm x_i$. This term dominates near the origin, forcing the density of eigenvalues $\rho(x)$ to vanish as $x \to 0$. For $x \ll N^{-1/2}$, the density is a linear function, while for larger $x$ it is well approximated by the semicircle distribution~\cite{Altland:1997zz}. In more detail, the approximate distribution of neutrino Majorana masses is
\be\label{eq:AZ-neutrinos}
\renewcommand{\arraystretch}{1.5}
\rho(m) \sim \left\lbrace
\begin{matrix}
\vspace{3mm}
N\, |m| \; &\quad\text{for}&\; |m| \ll \frac{1}{N}  \,, \\
N\sqrt{4-m^2}\; &\quad\text{for}&\;  \frac{1}{N}\ll |m| \le 2 \,,
\end{matrix} \right. 
\ee
where we have dropped numerical constants and the range of the distribution (here set to $|m|=2$) is arbitrary. Exact formulas for the eigenvalue density and higher correlation functions can be obtained in terms of Laguerre polynomials~\cite{Slevin}.

\begin{figure}[t!]
\begin{center}
\includegraphics[width=0.4\textwidth]{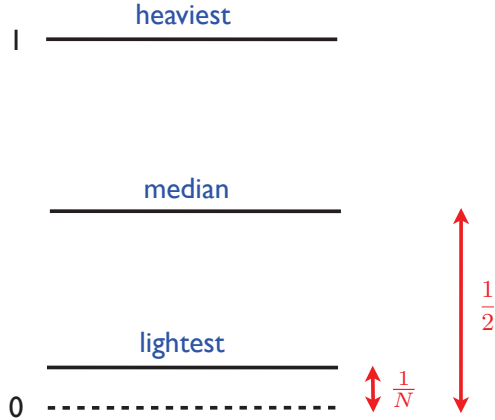}
\caption{ Qualitative spectrum for the complex symmetric case with a large $N$. The location of the lightest, median and heaviest eigenvalues are shown from bottom to top.}
\label{fig:complex-spec}
\end{center}
\end{figure}

Let us now explore the phenomenological consequences of these results. In the $N=3$ case we are interested in the spectrum of masses of the lightest, intermediate and heaviest neutrino. This is generalized to $N>3$ by looking at the expectation values of the lightest eigenvalue $m_1$, the median $m_{(N+1)/2}$ (or $m_{N/2}$ depending on whether $N$ is even or odd), and the heaviest $m_N$. Given the eigenvalue density $\rho(m)$, these expectation values are obtained from
\be\label{eq:mk-def}
k = \int_0^{m_k} dm\,\rho(m)\,.
\ee
For eigenvalue distributions that are approximately constant near the origin (such as the semicircle distribution of the GUE), $m_1/m_N \sim 1/\rho(0)$, which is the spacing of levels at small $m$. For complex Majorana neutrinos this estimate does not apply directly, because the distribution Eq.~(\ref{eq:AZ-neutrinos}) vanishes at the origin. However, the more precise Eq.~(\ref{eq:mk-def}) shows that the level spacing is parametrically of the same order as that of the semicircle law. Similarly, the median is found to be $\sim 1/2$. 

In summary, we have
\be
\frac{m_1}{m_N} \sim \frac{1}{N}\;,\qquad \;\frac{m_{N/2}}{m_N} \sim \frac{1}{2}\,.
\ee
We have also checked these results numerically, finding the same type of behavior. The neutrino mixing angles and phases are encoded in the unitary matrix $U$ of Eq.~(\ref{eq:Mnu-SVD}). At large $N$ these variables are distributed according to the Haar measure. For $N \to 3$ there are small deviations depending on whether the scan over neutrino matrices is done over a hypercube or a hypersphere $\tr (M^\dag M) \le 1$. The phenomenological predictions of this measure were studied in detail in~\cite{Hall:1999sn,Haba:2000be}.

This type of parametric spectrum is not the one suggested by experimental data (reviewed in \S\ref{sec:application}), at least in the limit where large $N$ results give a good approximation to the $N=3$ case. For this reason, we next turn to analyze the statistical properties of masses generated by the Seesaw mechanism.

\section{New hierarchies and ensemble from Seesaw neutrinos}
\label{sec:see-saw}

We now focus on the more interesting and nontrivial Seesaw models, which have been used to explain the smallness of the neutrino masses with respect to the electroweak scale. We will find that starting from random masses for the light active and heavy singlet neutrinos, the Seesaw mechanism leads to hierarchies of masses that are parametrically different from those obtained in \S \ref{sec:Majorana}. Furthermore, this mechanism will motivate the definition of a new type of random matrix ensemble, the ``Seesaw ensemble'', whose interesting mathematical properties we will begin to explore here.

\subsection{Random Seesaw neutrinos}
\label{subsec:random-seesaw}

Let us consider the scenario with $N$ active (left-handed) neutrinos, and $N$ singlet (right-handed) neutrinos which have heavy Majorana masses. The case with different numbers $N_L$ and $N_R$ of left- and right-handed neutrinos also appears to be interesting, offering the possibility of taking the large $N_R$ limit while keeping $N_L$ fixed. We will comment more on this possibility in \S \ref{sec:conclusion}. The neutrino mass matrix is a $2N\times 2N$ matrix of the form
\bea\label{eq:Mbig}
\left(
\renewcommand{\arraystretch}{1.5}
\begin{array}{cc}
0 &  M_D\\
M_D^T & M_R 
\end{array}
\right)
\eea
where $M_R$ is the symmetric complex Majorana mass matrix for the right-handed neutrinos, and $M_D$ is
the complex Dirac mass matrix between the left-handed and right-handed neutrinos. For $||M_R|| \gg ||M_D||$ the heavy SM singlet neutrinos can be integrated out to obtain the $N \times N$ matrix for the active neutrinos, 
\be
M_\nu\,=\,M_D\,M_R^{-1}\,M_D^T\,.
\label{eq:seesaw1}
\ee
This is the matrix that will be the object of our study.

Before proceeding, it is necessary to point out that the overall scales for $M_D$ and $M_R$ will not be addressed in the paper. We only assume that $||M_D|| \ll ||M_R||$, so that the approximation Eq.~(\ref{eq:seesaw1}) is valid and the Seesaw mechanism can be used to explain the smallness of the neutrino mass scale. 

We take $M_D$ and $M_R$ as random matrices and study the statistical properties of the eigenvalues of the matrix in Eq.~(\ref{eq:seesaw1}). Starting from these ensembles and using Eq.~(\ref{eq:seesaw1}), it should be possible to determine the spectrum and correlation functions for the eigenvalues of $M_\nu$. However, such a first-principle derivation turns out to be very involved, requiring the diagonalization of $M_\nu$ and the Jacobian matrix to rewrite the distributions for $M_D$ or $M_R$ in terms of the eigenvalues of $M_\nu$. Instead, we will consider an alternative strategy based on an ansatz for the eigenvalue distribution $\rho(m_\nu)$ for $M_\nu$, which will be verified numerically. Then we will apply the Coulomb gas picture to write down an action for the eigenvalues $m_\nu^i$, that can be used to derive the correlation functions of the system.

The first step is to diagonalize $M_D$ and $M_R$ separately, rewriting $M_\nu$ as
\be
\label{eq:seesaw2}
M_\nu =
\left(
\begin{matrix}
m_1 &  &  & \\
 & m_2 & & \\
 &   & \cdots & & \\
 & &       & m_N 
\end{matrix}
\right)
U_{\rm rel}
\left(
\begin{matrix}
M_1^{-1} &  &  & \\
 & M_2^{-1} & & \\
 &  & \cdots  &  \\
 &  &         &  M_N^{-1} \\
\end{matrix}
\right)
U_{\rm rel}^{T}
\left(
\begin{matrix}
m_1 &  &  & \\
 & m_2 & & \\
 &   & \cdots & & \\
 & &       & m_N 
\end{matrix}
\right)\,.
\ee
Here, $U_{\rm rel}$ is a unitary matrix that comes from the relative diagonalization of $M_D$ and $M_R$ and $m_i$ ($M_i$) are the eigenvalues of $M_D$ ($M_R$). The eigenvalues of $M_\nu$ will be denoted by $m_\nu^i$. In order to develop intuition on the spectrum of $M_\nu$, let us approximate the eigenvalues of $M_D$ and $M_R$ by
\beqa
m_i \approx \frac{i}{N}\,m \, \qquad \mbox{and}\qquad M_i \approx \frac{i}{N}\,M \,.
\eeqa
This is a good approximation away from the edges of the semicircle distributions.

We will find that the distribution of $M_\nu$ can be described in terms of an action Eq.~(\ref{eq:Hgeneral}), with a confining potential $V(m_\nu)$ and a logarithmic Coulomb repulsion. Let us now consider the effects from eigenvalue repulsion,
\beqa
H_{\rm repul} = - \sum_{i < j} \log \left( m^{j\,2}_\nu - m^{i\,2}_\nu \right) \,. 
\eeqa
The matrix $U_{\rm rel}$ rotates the relative orientation of $M_i$ and $m_j$ for the final eigenvalues. Although there are many possible relative orientations, we choose two points for illustration purpose. The first case is to have
\beqa
&& \hspace{-3.0cm} \mbox{Case A}:  \hspace{2.9cm} m^i_{\nu\,A} = \frac{m_i^2}{M_i} = \frac{m^2}{M}\,\frac{i}{N} \,,
\eeqa
which has $U_{\rm rel} = \mathbb{I}_N$. The second example is
\beqa
&&\mbox{Case B}:  \hspace{3cm} m^i_{\nu\,B} = \frac{m_i^2}{M_{N+1-i}} = \frac{m^2}{M}\,\frac{i^2}{N(N+1-i)} \,,
\label{eq:seesaw-mnuB}
\eeqa
with $U_{\rm rel}$ as a permutation matrix that interchanges $i\leftrightarrow N+1-i$ in $M_R$. In other words, this case corresponds to having the largest suppression for the smallest $M_\nu$ eigenvalue. 

\begin{figure}[t!]
\begin{center}
\includegraphics[width=0.4\textwidth]{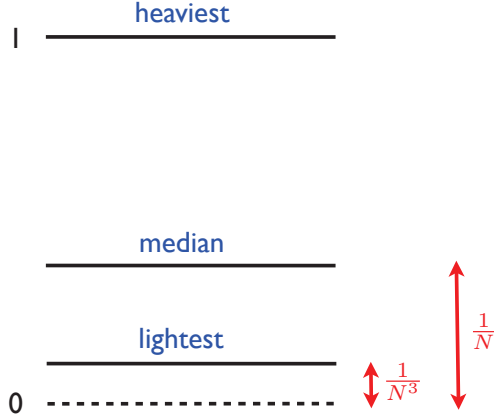}
\caption{Qualitative spectrum for the Seesaw case with a large $N$. The location of the lightest, median and heaviest eigenvalues are shown from bottom to top.}
\label{fig:seesaw-spec}
\end{center}
\end{figure}

Comparing the repulsion Hamiltonian for these two choices of orientations, we have
\beqa
H_{\rm repul, A} - H_{\rm repul, B}
= - \sum_{i<j} \left\{   
\log{(j^2 - i^2)} - \log{\left[ \left(\frac{j^2}{N+1-j}\right)^2 - \left(\frac{i^2}{N+1-i}\right)^2  \right]}
\right\} > 0 \,,
\eeqa
which shows that the case B has a smaller energy from contribution of the Coulomb repulsion. Concentrating then on the eigenvalue distribution of the case B (we drop the label ``$B$" from now on), we have, for the smallest eigenvalue and the median,
\beqa\label{eq:seesaw-small-median}
\frac{m^1_{\nu}}{m^N_{\nu}} = \frac{1}{N^3} \,,\qquad\qquad
\frac{m^{N/2}_{\nu}}{m^N_{\nu}} = \frac{1}{2\,N} \,.
\eeqa
This is shown qualitatively in the spectrum in Fig.~\ref{fig:seesaw-spec}. Based on numerical results, we will now argue that the case B gives a very good approximation to the full eigenvalue distribution.

The spectrum of $M_\nu$ is studied numerically as follows. Treating all elements of $M_D$ and $M_R$ as complex numbers, we choose random numbers for both the real and imaginary parts of the elements of $M_D$ within $(-1, 1)$ and $(-10^{10}, 10^{10})$ for $M_R$, which incorporates the two hierarchic scales in the Seesaw mechanism. We calculate $M_\nu$ and diagonalize it. This procedure is repeated several thousand times to calculate the eigenvalue distribution. Let us now focus on the smallest eigenvalue and the median, and compare with the predictions Eq.~(\ref{eq:seesaw-small-median}). The full distribution will be studied below.

\begin{figure}[ht!]
\begin{center}
\includegraphics[width=0.45\textwidth]{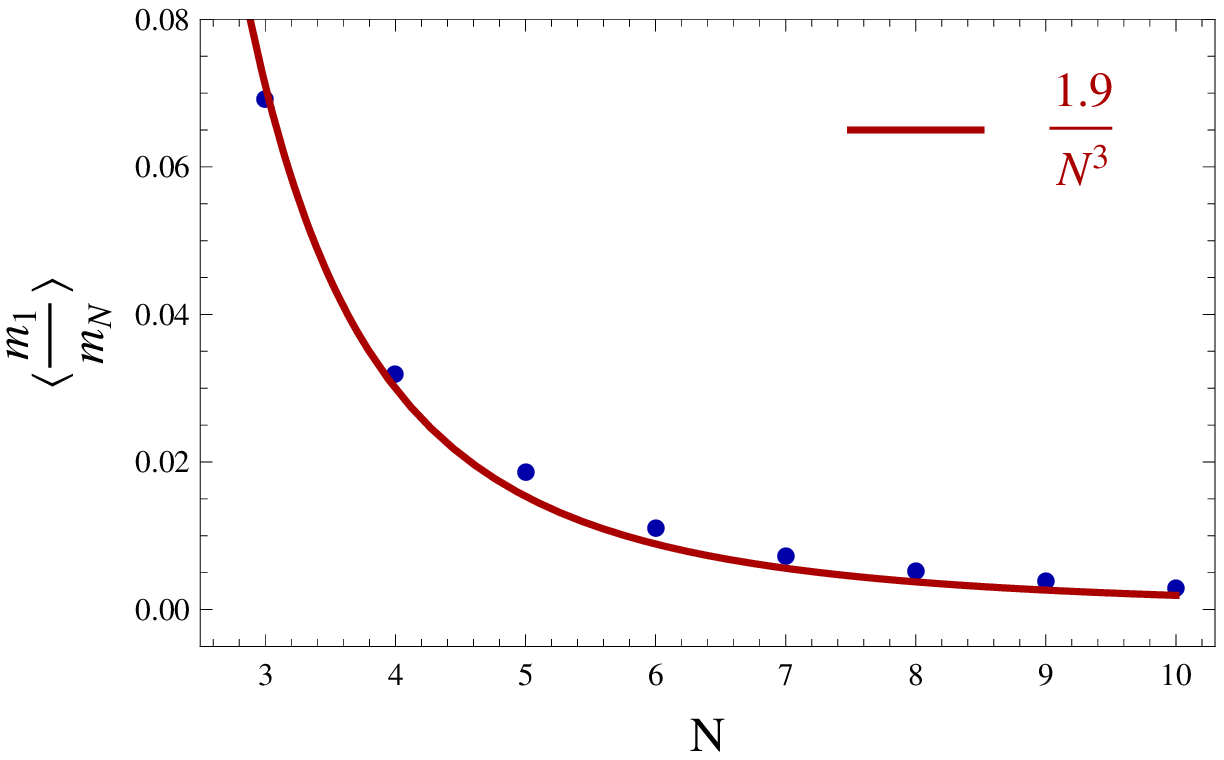} \hspace{5mm}
\includegraphics[width=0.45\textwidth]{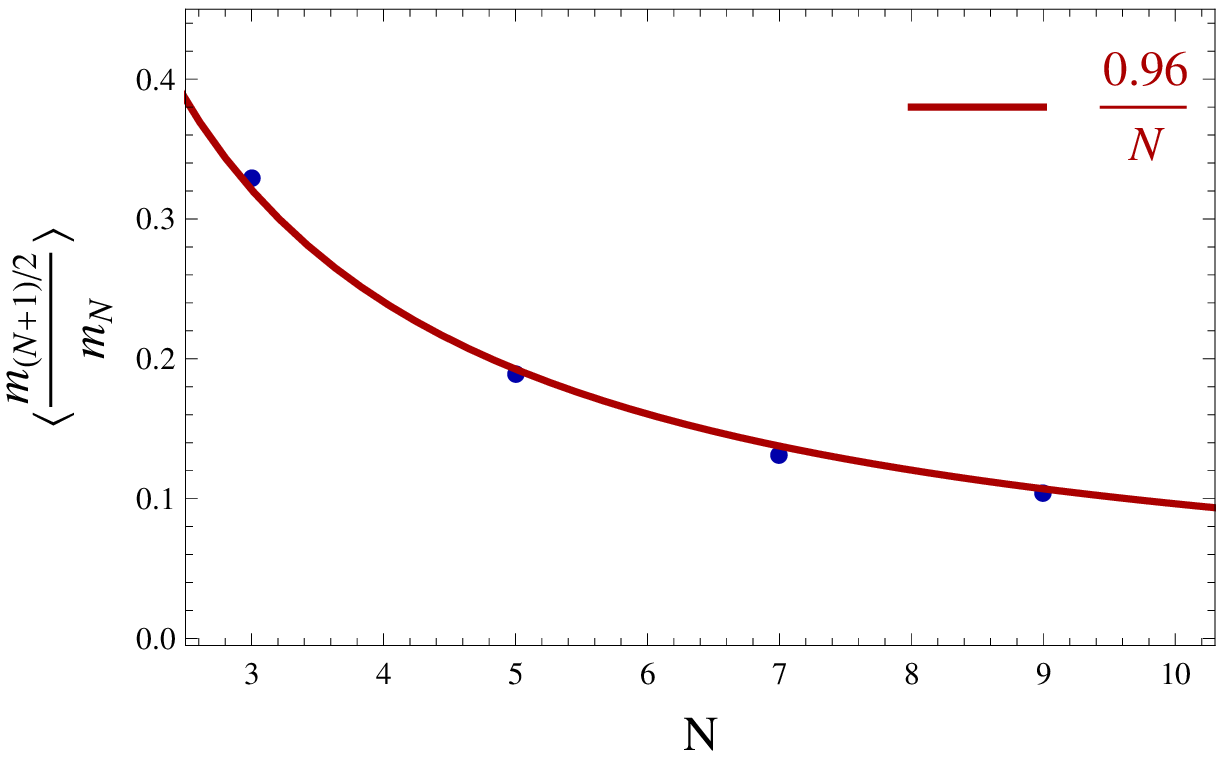}
\caption{Left panel: the blue circle points are the ratios of the smallest eigenvalue over the largest eigenvalues as a function of the matrix rank for the Seesaw case. The red continuous line is the fitted result using a function proportional to $N^{-3}$. Right panel: the same as the left but for the median eigenvalue over the largest eigenvalue. The fitted function is chosen to be $N^{-1}$.}
\label{fig:seesaw-ratio}
\end{center}
\end{figure}

The numerical results for the smallest eigenvalue as a function of $N$
 are shown in the blue circles of the left panel of Fig.~\ref{fig:seesaw-ratio}, together with a fitted distribution using a function proportional to $N^{-3}$. One can see that the $N^{-3}$ is indeed a good description of the large $N$ behavior and also of the $N=3$ case. The right panel of Fig.~\ref{fig:seesaw-ratio} shows the ratio of the median eigenvalue over the largest eigenvalue. (For simplicity, only odd $N$ numbers are shown in this plot.) The numerical results are well fitted by a $N^{-1}$ function. Therefore, we find a good agreement between Eq.~(\ref{eq:seesaw-small-median}) and the numerical eigenvalues; this provides nontrivial evidence that the distribution Eq.~(\ref{eq:seesaw-mnuB}) is a good approximation. Further properties of the eigenvalue distribution will be analyzed shortly.

\begin{figure}[ht!]
\begin{center}
\includegraphics[width=0.48\textwidth]{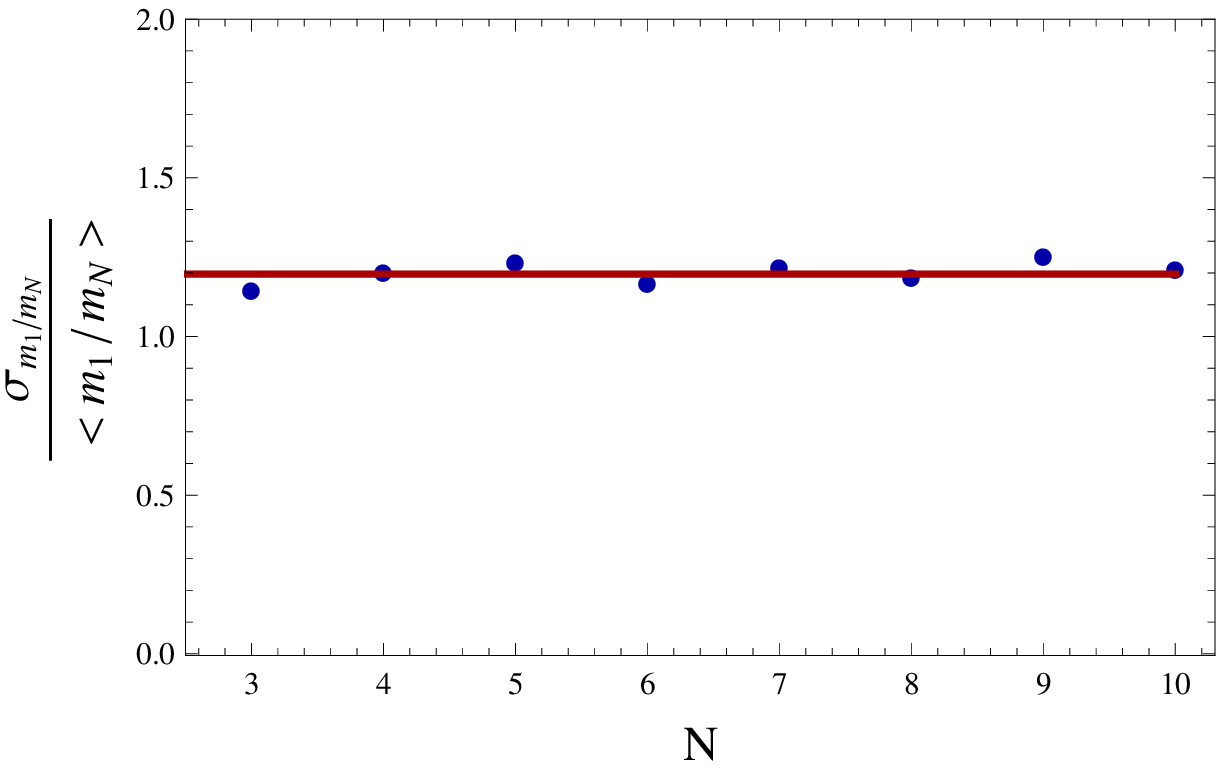} 
\hspace{3mm}
\includegraphics[width=0.48\textwidth]{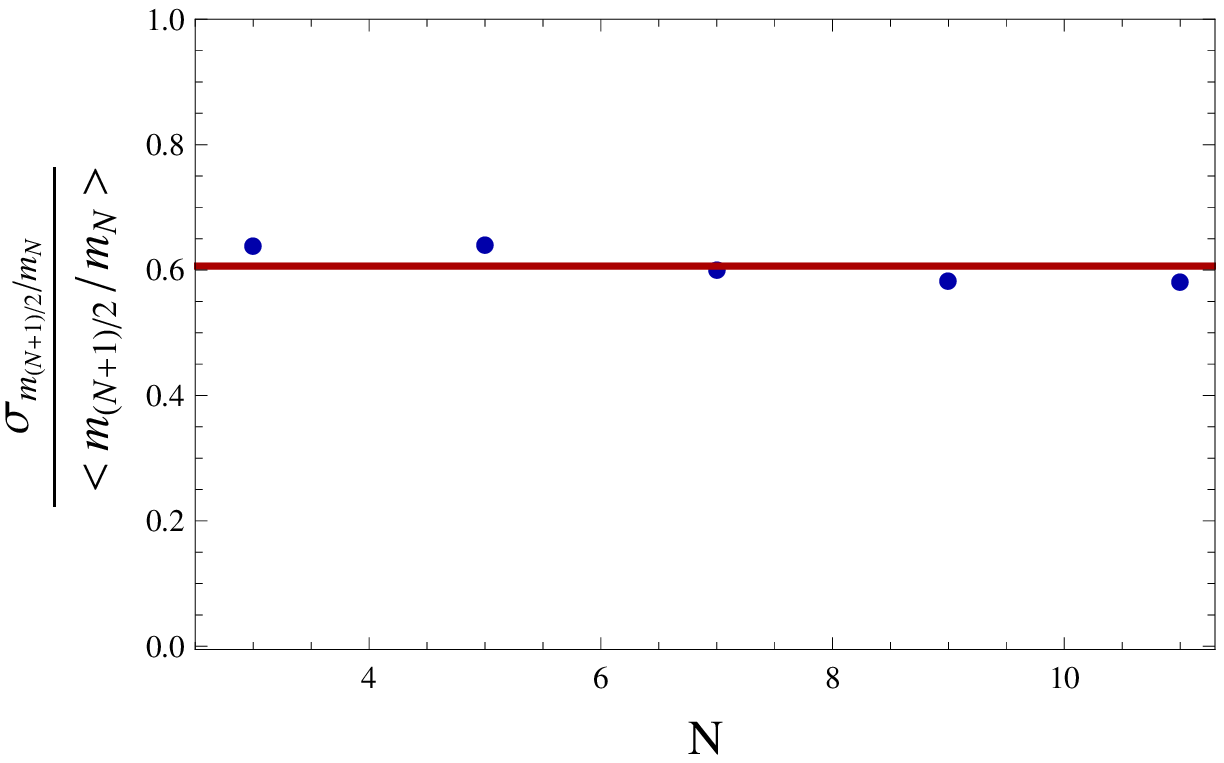} 
\caption{The ratios of the variance over the expectation value as a function of $N$.}
\label{fig:seesaw-sigma}
\end{center}
\end{figure}

Since we are doing statistical analysis for the expectation values of the eigenvalue ratios, it also important to compute the standard deviation of these observables as a function of $N$. Defining 
\beqa
\sigma_{m_i/m_j} \equiv \sqrt{ \left\langle  \frac{m^2_i}{m^2_j}   \right\rangle   -  
\left\langle  \frac{m_i}{m_j}  \right\rangle^2    }   \,,
\eeqa
the large $N$ dependence of $\sigma_{m_i/m_j}$ turns out to be the same as that of $m_i/m_j$. This is illustrated in Fig.~\ref{fig:seesaw-sigma}, which shows the numerical ratios of the variance over the expectation value. This implies that our large $N$ analysis can not capture detailed order one coefficients in front of the $N$ dependence.

Comparing Fig.~\ref{fig:complex-spec} and Fig.~\ref{fig:seesaw-spec} shows that the Seesaw mechanism generates a larger hierarchical spectrum than the one obtained from random Majorana masses. For a fixed heaviest eigenvalue, random Majorana neutrinos have a lightest eigenvalue suppressed by $1/N$, while in the Seesaw case the suppression is by $1/N^3$. The behavior of the median is also parametrically different as a function of $N$. These statistical properties are then very sensitive to the mass mechanism, providing predictions that distinguish the two scenarios.
In \S\ref{sec:application} we will compare those two spectra to the experimentally observed values, and conclude that the Seesaw scenario is preferred.

\subsection{The Seesaw Ensemble}
\label{subsec:ensemble}

So far we have found that the Seesaw mechanism provides an interesting way of generating additional hierarchies out of randomness. Starting from a small input parameter of $1/N$, it produces a parametrically smaller $1/N^3$ suppression. In terms of the original masses in $M_R$ and $M_D$, with $||M_R|| \gg ||M_D||$, this is the statement that the expectation values are dominated by the relative orientation Eq.~(\ref{eq:seesaw-mnuB}). This effect may also have applications in other systems. For this reason, we now study in more detail the ``Seesaw ensemble'' by itself, independently of motivations from neutrino physics. As far as we know, this provides a new type of ensemble beyond the ones studied in the literature (see e.g.~\cite{Stephanov:2005ff} for a recent review).

We define the Seesaw ensemble as the set of $N \times N$ matrices of the form
\beq\label{eq:Mnu2}
M_\nu = M_D M_R^{-1} M_D^T\,,
\eeq
where $M_D$ and $M_R$ are complex random $N \times N$ matrices, and $M_R$ is symmetric. Recalling the discussion in \S\ref{sec:Majorana}, $M_R$ belongs to the AZ-CI ensemble.\footnote{In particular, the eigenvalue density vanishes at the origin.} On the other hand, the probability distribution for the Dirac matrices $M_D$ is given by~\cite{Haba:2000be}
\beq
P(M_D) dM_D \propto \prod_k m_k dm_k\,\prod_{i<j} (m_i^2 -m_j^2)^2\, e^{-\frac{N}{4} \sum_i m_i^2}\,.
\eeq
This is in the same class as the unitary ensemble that appears in QCD, with no flavors and a nonchiral Dirac spectrum.

Ideally we would want to derive the probability distribution for $M_\nu$ starting from those of $M_D$ and $M_R$. This is, however, a very complicated task. Instead, we now explore the consequences of the ansatz Eq.~(\ref{eq:seesaw-mnuB}) for the spectrum of eigenvalues of $M_\nu$, which we found to be in good agreement with numerical results. Our analysis of the Seesaw ensemble will rely heavily on the generalized Coulomb gas picture described in \S \ref{subsec:coulomb}.

From Eqs.~(\ref{eq:mk-def}) and (\ref{eq:seesaw-mnuB}), it is not difficult to derive the continuous spectral density for the Seesaw ensemble,
\beqa
\rho^N_{\rm{Seesaw}}(x) =\frac{1}{2} \frac{N^2}{N+1} \frac{ x N^2  + 2 N + 2 - N\,\sqrt{x^2 N^2  + 4 x N  + 4 x}}{\sqrt{x^2 N^2  + 4 x N  + 4 x}} \,,
\label{eq:rho-seesaw}
\eeqa
where the normalization is such that $\int^\infty_0 dx \rho(x) = N$. For very small and large values of $x$, the density behaves as
\beqa
x\rightarrow 0: \quad \rho^N_{\rm{Seesaw}}(x) \rightarrow \frac{N^2}{2\sqrt{N+1} \sqrt{x}}\,, \qquad\qquad
x\rightarrow \infty: \quad \rho^N_{\rm{Seesaw}}(x) \rightarrow \frac{N+1}{N\,x^2} \,.
\eeqa
Both limits reveal interesting properties of the ensemble. Note that the spectral density is divergent when $x\rightarrow 0$, but $\rho(x)$ is still integrable. Furthermore, $\rho(x)$ does not have compact support --there is no upper bound on eigenvalues for the Seesaw ensemble. This comes from the inverse of $M_R$ in Eq.~(\ref{eq:Mnu2}), so that it is always possible to have eigenvalues much larger than the light scale $m^2/M$. These properties are in sharp contrast with the GUE or AZ-CI ensembles that we discussed before, for which $\rho(x)$ has compact support and is finite as $x \to 0$.

\begin{figure}[t!]
\begin{center}
\includegraphics[width=0.48\textwidth]{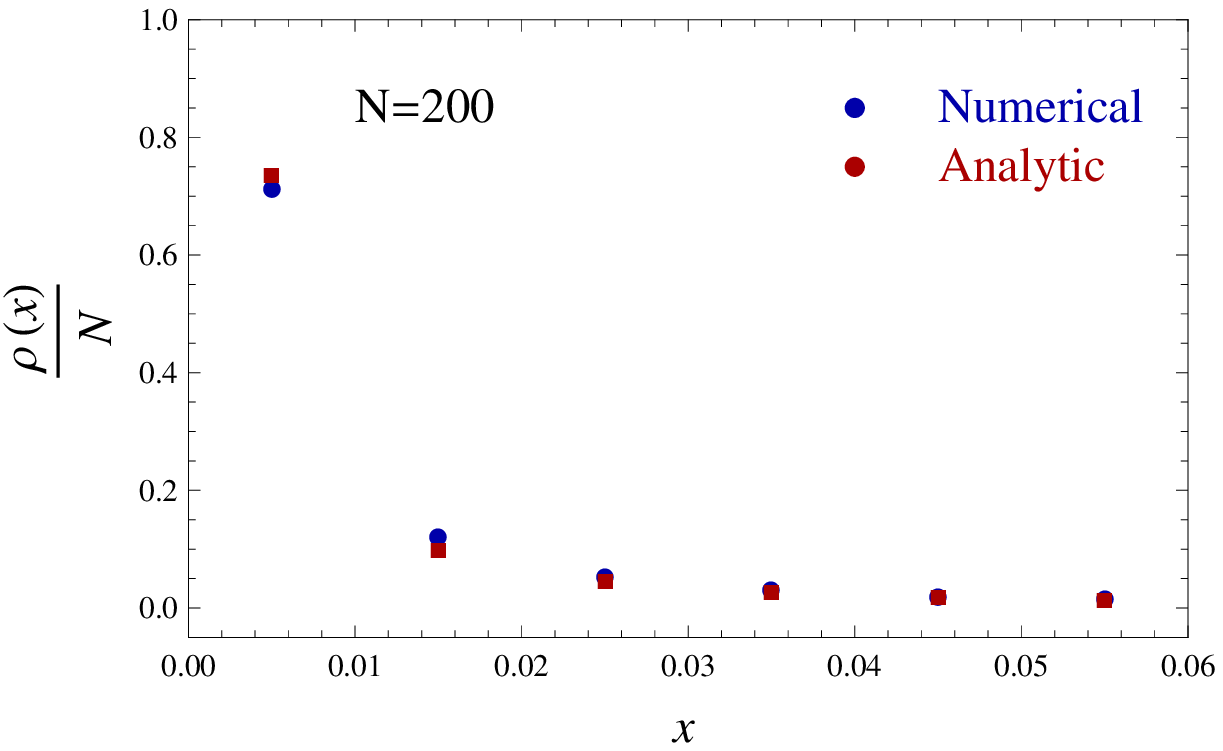} 
\hspace{3mm}
\includegraphics[width=0.48\textwidth]{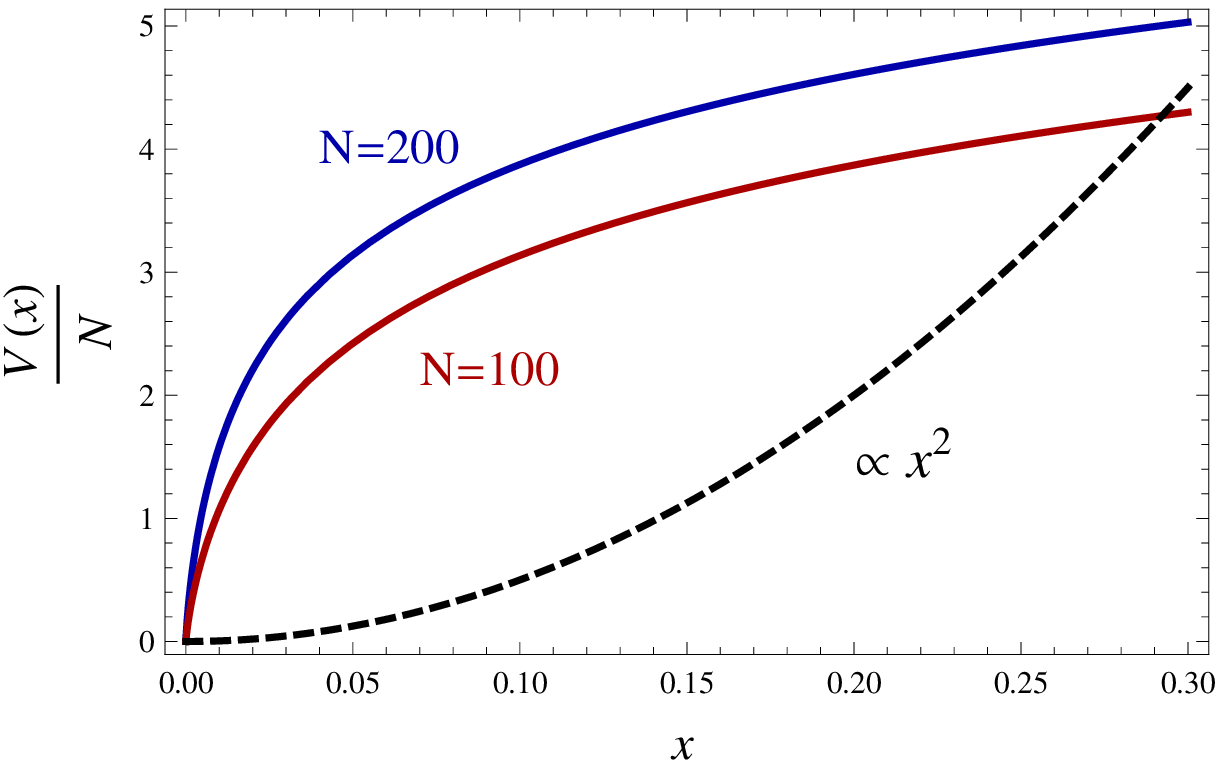} 
\caption{Left panel: the normalized spectral density as a function of eigenvalues for the Seesaw ensemble. The points for the analytic results are calculated by integrating $\rho(x)$ for the bin size. Right panel: the red (solid) and blue (solid) lines are the effective confinement potentials for $N=100$ and $N=200$, respectively, in the Coulomb gas picture. The black (dashed) line is the harmonic potential for the GUE.}
\label{fig:seesaw-spectra}
\end{center}
\end{figure}

In order to verify our analysis so far, let us compare the spectral density Eq.~(\ref{eq:rho-seesaw}) with numerical results. This is done in the left panel of Fig.~\ref{fig:seesaw-spectra}, which shows both the numerically averaged density of eigenvalues for $N=200$ and the analytic prediction. One can see a good agreement between those two distributions. 

Let us now develop the Coulomb gas model for the Seesaw ensemble. As we reviewed in \S\ref{sec:RMT}, this representation is given by a system of particles moving along a line in two dimensions, subject to a Coulomb repulsion and a confining potential $V(x)$. We already discussed the effects from repulsion, and now we will determine $V(x)$. Plugging the eigenvalue density into Eq.~(\ref{eq:Vrho}) obtains
\beq
V^N_{\rm Seesaw}(x) = \frac{N^2}{2(N+1)} \left[ N x \log{\left( \frac{N+1}{N^2\,x} \right)}
 + 2 \sqrt{ N^2 x^2 +4 N x +4 x} \,\tanh^{-1} \left( \sqrt{\frac{N^2 x}{ N^2 x + 4 N +4 }} \right)  \right] \,.
\eeq
Mapping back to the matrix eigenvalues, this means that the probability distribution for Seesaw matrices is
\beq
P(M_\nu) d M_\nu \propto \prod_k dm_k \prod_{i<j} \left|m_i^2-m_j^2\right|\, e^{-N\,V(m_i^2)}\,.
\eeq

In the limits of small and large values of $x$, we have
\beqa
x\rightarrow 0: \; V^N_{\rm Seesaw}(x) \rightarrow \frac{N^3 x \left[ \log{\left( \frac{N+1}{N^2 x} \right) } +2  \right]}{2(N+1)}
\;,\;
x\rightarrow \infty: \; V^N_{\rm Seesaw}(x) \rightarrow 
N - N \log{\left[  \frac{(N+1)}{N^2 x}  \right]}
\,.
\eeqa
Thus, the confining potential $V^N_{\rm Seesaw}(x)$ is continuous at $x=0$, but its first derivative diverges there. This is much steeper than the harmonic potential for the GUE, explaining why the eigenvalue density is parametrically larger near the origin.\footnote{See for instance~\cite{qRandom} for different potentials in other q-random matrix ensembles.} The Coulomb gas picture hence provides a physical way of understanding the steeply decreasing behavior of $\rho(x)$ as $x$ increases.

Although we will not study higher correlators in detail, let us mention that given $V(x)$, the correlation functions for the Seesaw ensemble can be obtained as functions of polynomials that are orthogonal with respect to $e^{-V(x)}$. They are of the form Eq.~(\ref{eq:Rk}) in terms of these new orthogonal polynomials.

\section{Application to neutrino physics}
\label{sec:application}

Having analyzed the statistical properties of neutrinos for general $N$, let us now focus more specifically on neutrino physics, comparing our results (extrapolated to $N=3$) with the experimental data.

From all neutrino experimental results including solar, atmospheric, reactor and accelerator experiments, the global analysis~\cite{GonzalezGarcia:2012sz} obtains the following numbers for the neutrino mass squared differences and mixing angles, defined in the PDG~\cite{Beringer:1900zz}:
\beqa
&& \Delta m^2_{21} = (7.50\pm 0.185)\times 10^{-5}~\mbox{eV}^2 \,,\\
&&\Delta m^2_{31}\,[\mbox{N}] = (2.47^{+0.069}_{-0.067})\times 10^{-3}~\mbox{eV}^2\,, \qquad
\Delta m^2_{32}\,[\mbox{I}] = (-2.43^{+0.042}_{-0.065})\times 10^{-3}~\mbox{eV}^2\,,
\\
&& \sin^2{\theta_{12} } = 0.30\pm 0.013 \,, \qquad \sin^2{\theta_{23}}=0.41^{+0.037}_{-0.025} \oplus 0.59^{+0.021}_{-0.022}\,, \\
&& \sin^2{\theta_{13}} = 0.023\pm 0.0023 \,, \qquad \delta_{\rm CP} = 300^{\circ\,+66}_{\,\,-138} \,.
\eeqa
Here, $\Delta m^2_{ij} \equiv m^2_i - m^2_j$; ``N" means the normal ordering for the neutrino spectrum such that $\Delta m^2_{21}\ll (\Delta m^2_{32} \approx \Delta m^2_{31} > 0)$; ``I" means the inverted ordering for the neutrino spectrum such that $\Delta m_{21}^2\ll -(\Delta m^2_{31} \approx \Delta m^2_{32} < 0)$. We take their fitted results for treating the reactor fluxes as free parameters and including the results for the reactor experiments with $L \lesssim 100$~m. As emphasized in~\cite{GonzalezGarcia:2012sz}, although a zero Dirac CP violation phase $\delta_{\rm CP}$ is referred, the significance is less than 1.5$\sigma$ for the normal ordering and 1.75$\sigma$ for the inverted ordering. The absolute scale of neutrino masses is currently unknown and bounded from above as  $\sum m_\nu \leq 0.28$~eV at the 95\% confidence level~\cite{Thomas:2009ae}, assuming a flat $\Lambda$CDM cosmology. 

We should emphasize that our analysis does not explain the overall scale of neutrino masses, but only ratios of masses. Let us concentrate first on the ratio $\Delta m^2_{21}/\Delta m^2_{31} \approx 0.03$. For complex Majorana masses without the Seesaw mechanism (see Fig.~\ref{fig:complex-spec}), the prediction is $\Delta m^2_{21}/\Delta m^2_{31} = {\cal O} (1)$ at large $N$, which does not provide a good fit to the experimental value. On the other hand, the prediction for Seesaw masses (see Fig.~\ref{fig:seesaw-spec}) is $\Delta m^2_{21}/\Delta m^2_{31} = {\cal O} (1/N^2) = {\cal O}{(0.1)}$ for $N=3$, much closer to the experimental result. (Recall that these predictions have order one uncertainties.) Therefore, assuming random masses, the Seesaw mechanism is favored by data.

Focusing then on the Seesaw mechanism, let us make the following observations based 
on the spectrum in Fig.~\ref{fig:seesaw-spec}. First, the neutrino spectrum is a normal hierarchical spectrum. The general neutrino mass scale is determined by the heaviest mass, $m_3\approx\sqrt{\Delta m^2_{32}} \approx 0.05$~eV, which could be tested from the Cosmic Microwave Background  by the Planck satellite or the KATRIN experiment~\cite{Aseev:2011dq} in the future. The prediction for the effective Majorana mass $|\langle m \rangle|$, which may be measured from neutrinoless double beta decay experiments, is in the region of $(2.3\times 10^{-4}, 5.0\times 10^{-3})$~eV for the normal hierarchical spectrum~\cite{Beringer:1900zz}. Secondly, the lightest neutrino mass is predicted to be $m_3/N^3 = m_3/3^3 \approx 0.002$~eV, which is unlikely to be tested in the near future unless one can come up a new experimental method to measure the individual neutrino mass more precisely.

Finally, let us discuss the basic properties of the mixing matrix for this case. As mentioned in~\cite{Haba:2000be}, the basis-independence assumption predicts that the PMNS matrix follows the Haar measure of the Lie group. While in principle the corresponding measure for the Seesaw ensemble can be obtained starting from the Haar measure distributions for $M_R$ and $M_D$, for our purpose it is enough to evaluate numerically the expectation values of the elements of the matrix $U$ that diagonalizes $M_\nu^\dag M_\nu$.
\begin{figure}[t!]
\begin{center}
\includegraphics[width=0.48\textwidth]{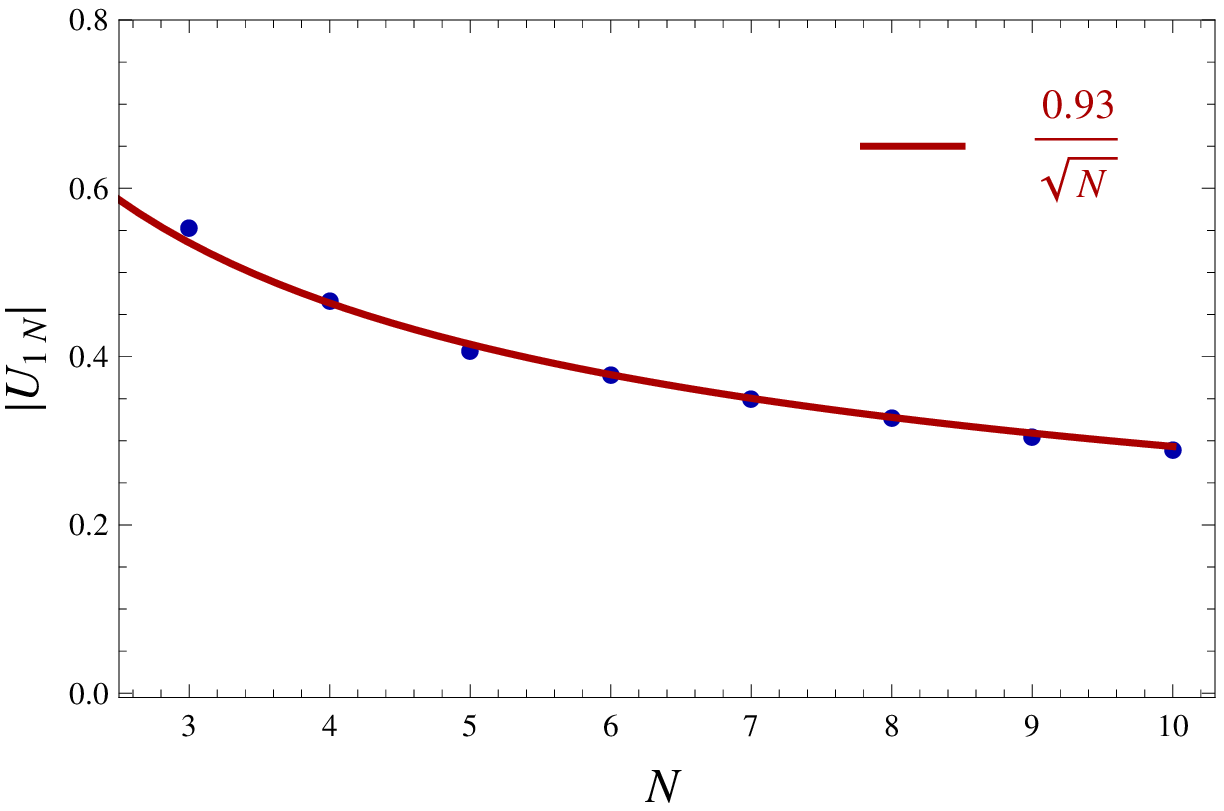} 
\hspace{3mm}
\includegraphics[width=0.48\textwidth]{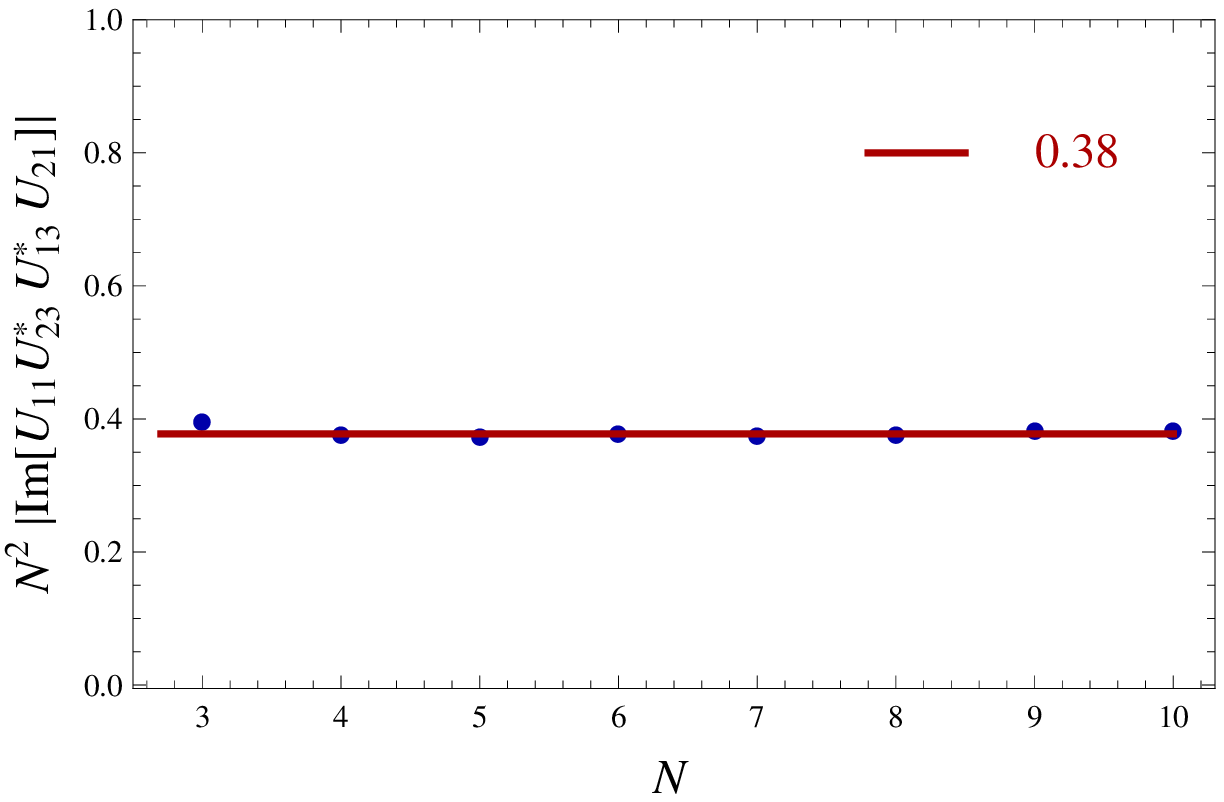} 
\caption{Left panel: the absolute value of the off-diagonal element $U_{1N}$ as a function of $N$. Right panel: the absolute value of the Jarlskog invariant multiplying $N^2$, which denotes the size of the physical phases in the unitary matrix, as a function of $N$.}
\label{fig:seesaw-angle-phase}
\end{center}
\end{figure}
The left panel of Fig.~\ref{fig:seesaw-angle-phase} shows the behavior of the absolute value of the off-diagonal entry $U_{1N}$ as a function of $N$, which is proportional to $1/\sqrt{N}$. This is just what we expect based on the normalization of eigenvectors. For the phases in the rotation matrix, we use the generalized Jarlskog invariants to study the rephasing invariant CP violation~\cite{Jarlskog:1985ht, Bjorken:1987tr}. In the right panel of Fig.~\ref{fig:seesaw-angle-phase}, we show the distribution of $N^2 \left |\mbox{Im} [U_{11}U^*_{23} U^*_{13} U_{21} ]\right |$ as a function of $N$. The overall factor $N^2$ cancels the $N$ dependence of the absolute values of the elements. We can see from this panel that the phase part of this Jarlskog invariant is order $1$ and independent of $N$.

Translating these results into mixing angles and phases, our prediction for $\sin{\theta_{13}}$ is order $1/N = 1/3$, not far from the measured values $\sin{\theta_{13}}\approx 0.15$. For the CP violation phases, we anticipate all the phases in the leptonic mixing matrix in the weak charged current interactions to be order $1$. The predictions of our analysis are summarized in Table~\ref{tab:seesaw-prediction}.
\begin{table}[t!]\small
\vspace*{4mm}
\renewcommand{\arraystretch}{2.5}
\centerline{
\begin{tabular}{c|c}
\hline \hline
Spectrum & Normal hierarchical  \\  \hline 
Lightest neutrino mass & $\sqrt{\Delta m^2_{32} }\times {\cal O} \left[(N=3)^{-3}\right] \approx 0.002$~eV   \\ \hline
$\sin{\theta_{13}}$ & ${\cal O}\left[(N=3)^{-1}\right] \approx 0.3$  \\
\hline
$\delta_{\rm CP}$  & ${\cal O} (1)$ \\ \hline \hline
\end{tabular}
}
\caption{The large $N$ predictions for the neutrino properties for the three active neutrinos.}
\label{tab:seesaw-prediction}
\end{table}
%

\section{Conclusions and future directions}
\label{sec:conclusion}

In this work we proposed to treat the number $N=3$ of Standard Model generations as a large number, studying the consequences of this idea for neutrinos. Assuming anarchic neutrino masses, this becomes a problem in Random Matrix Theory. We analyzed the neutrino spectrum and mixing matrices analytically and numerically, and compared with experimental results. The statistical properties of these observables were found to be highly sensitive on the underlying mechanism for neutrino masses. In particular, the Seesaw mechanism leads to a $1/N^3$ hierarchy for the lightest neutrino, and has a spectrum that is qualitatively compatible with experimental data.

Already at this level, the connections between neutrino physics and RMT appear to be interesting. Based on the Seesaw mechanism, we defined a new type of random matrix ensemble, the ``Seesaw ensemble,'' which, as far as we know, has not been analyzed before in the literature. The properties of this ensemble are qualitatively different from those of Gaussian distributed matrices. We proposed an ansatz for the spectrum that was in good agreement with the numerical results. In the Coulomb gas picture, the confining potential for the Seesaw matrices is much steeper than in the Gaussian case, with a diverging first derivative at the origin.

It will be interesting to determine the correlation functions of the Seesaw ensemble starting from the distributions of the underlying matrices $M_D$ and $M_R$ in Eq.~(\ref{eq:Mnu2}). We hope that the analytic and numerical results reported here can shed light on this problem. There are also various possible extensions. First, one could consider directly the ensemble of matrices of the form Eq.~(\ref{eq:Mbig}), without restricting from the start to the limit $||M_R || \gg ||M_D||$. Furthermore, there is also a natural two-parameter generalization of this ensemble, where $M_R$ is a square $N_R \times N_R$ matrix, but $M_D$ is an $N_L \times N_R$ matrix. Physically, this corresponds to having $N_L$ active and $N_R$ singlet neutrinos~\cite{Gluza:2011nm}. One could then study the statistical properties at fixed $N_L$ and large $N_R$. More generally, the existence of the new parameter $N_R/N_L$ can lead to different large $N$ behaviors or multiscale effects. This is reminiscent of QCD with $N_f$ fundamental fermions, where taking $N_f/N_c \sim 1$ led to the Veneziano or topological limit~\cite{Veneziano:1976wm}.

The large $N$ limit can also be applied to neutrino physics with one or more sterile neutrinos or the SM quarks and charged leptons. The dramatically hierarchical pattern of quark masses and angles implies that the couplings in the Lagrangian cannot be randomly distributed --more structure is required to explain the flavor problem. Nevertheless, extra-dimension models such as~\cite{ArkaniHamed:1999dc} still allow for random coefficients, and it would be interesting to understand how the dynamics in the quark sector is modified by large $N$ statistical effects.

Finally, given the incredible range of applications of RMT, the Seesaw ensemble may occur in other types of systems. One would need some pattern of the form Eq.~(\ref{eq:Mnu2}), arising from the interplay of two hierarchical scales.

\subsection*{Acknowledgments}
We would like to thank Andr\'{e} de Gouv\^{e}a, Shamit Kachru, Michael Peskin and Gary Shiu for encouragement and useful discussions. YB is supported by the startup funding from the University of Wisconsin-Madison. SLAC is operated by Stanford University for the US Department of Energy under contract DE-AC02-76SF00515. This work was supported in part by the National Science Foundation under grant PHY-0756174 and by the Department of Energy under contract DE-AC03-76SF00515. YB also thanks the Aspen Center for Physics, under NSF Grant No. 1066293, where part of work was completed.

\providecommand{\href}[2]{#2}\begingroup\raggedright\endgroup

 \end{document}